\documentclass[twoside,fleqn]{article}
\usepackage{espcrc2}
\usepackage{epsfig}
\usepackage{amstex}
\usepackage{amssymb}
\newcommand{\psibar} {\bar{\psi}}

\newcommand{\mq} {\ensuremath{m_{\text{q}}}}

\newlength{\hepwidth}

\begin{document}

\title{\settowidth{\hepwidth}{hep-lat/9712012}
\hfill\parbox{1.1\hepwidth}{
{hep-lat/9712012}}\\
Tempered Fermions in the Hybrid Monte Carlo Algorithm
}

\author{ G. BOYD\thanks{Combined proceedings for Lattice 97, Edinburgh and the 
     International Workshop 'Lattice QCD on Parallel Computers',
     University of Tsukuba, Japan.}\address{ Centre for Computational Physics, 
University of Tsukuba, 
Tsukuba, Ibaraki 305, Japan. \hfill\\
email:boyd rccp.tsukuba.ac.jp 
}
}

\begin{abstract}

Parallel tempering simulates at many quark masses simultaneously, by changing
the mass during the simulation while remaining in equilibrium. The algorithm is
faster than pure HMC if more than one mass is needed, and works better the
smaller the smallest mass is.

\end{abstract}

\maketitle

\section{INTRODUCTION}
The standard algorithms used for full QCD are painfully slow.
Physically large objects, like instantons, pions
etc., may well decorrelate much more slowly than, say, the plaquette~\cite{us}.

Tempered algorithms~\cite{us,ST,st_biel} promote a parameter of the theory to a
dynamical variable that changes during the simulation, which has tremendous
potential for speeding up slow simulations. They have been successfully
implemented in $\beta$ for spin glasses, U(1) {\em etc}. For QCD at zero
temperature promoting $\beta$ does not help (although at high temperature,
around the chiral phase transition it probably will), but promoting the quark
mass, and allowing it to change during the simulation, does speed things up.

The mass is only changed if the configuration is simultaneously in the
equilibrium distribution of both masses. So tempering is always in equilibrium
and requires no re-weighting etc. afterwards.

The minimum  gain comes from running at heavier (faster) masses between independent
configurations. The maximum gain comes if the relevant auto-correlations 
are smaller for larger masses.

\section{TEMPERING}

The quark mass, $a\mq$ for staggered or $\kappa$ for Wilson fermions, becomes a
dynamic variable, and may take a different value for each trajectory of, say,
the hybrid Monte Carlo algorithm~\cite{hmca}. (Regard $m$ below as the
corresponding $\kappa$ for Wilson fermions.) The masses used belong to an
ordered set with $N_{m}$ elements, $[m_{\text{min}}, ... ,m_{\text{max}}]$.
The only requirement is that the action histograms of neighbouring masses
overlap.

There are two types of tempering, simulated and parallel tempering. 
The idea behind simulated tempering in QCD, investigated in~\cite{st_biel}, is
very simple, and as it is the basis for the parallel tempering investigated
here it will be described first.

In simulated tempering you add to the probability distribution a constant
$g_{i}$ for each mass $m_{i}$, which indicates roughly where the half-way point
between the action histograms of mass $m_{i}$ and $m_{i+1}$ is.  The original
QCD probability distribution $P(U,\phi)$ now becomes
\begin{equation*}
P(U,\phi,i) \propto \exp[-S(U,\phi,\beta,m_{i}) + g_{i}].
\end{equation*}
This distribution is simulated using your favourite algorithm for fixed quark
mass (eg., HMC here), combined with Metropolis steps to change from $m_i$ to
$m_{i\pm 1}$. The constants $g_{i}$ are only to enable the masses to change
both up and down, and do not affect the physics.

The hybrid Monte Carlo algorithm insures that the correct Gibbs distribution is
generated at each value of the mass, and the tempering Metropolis step insures
that the mass only changes if the configuration is part of the equilibrium
distribution of both masses.

The constants $g_{i}$ can be chosen freely, depending on what seems best for
the simulation. They do not affect the physics, only the frequency with which
each mass is visited. The $g_{i}$ can be fixed by choosing, for example, to
visit each mass with equal probability, $P(i)=1/N_{m}$. Then $g_{i} = -\ln
Z_{i}$, ie. the original free energy at fixed mass $m_{i}$. The choice is
arbitrary, though, and can be optimized for speed.

The simulation only needs $g_{i+1}-g_{i}$, and a good starting point is to take
the first two terms below:
\begin{equation*}
\Delta g = - \langle\psibar\psi\rangle V\delta m
                - \langle\chi\rangle V(\delta m)^{2}
                + O((\delta m)^{3})
\label{eq:delg}
\end{equation*}
where $\langle\psibar\psi\rangle$ and $\langle \chi\rangle$ are the chiral
condensate and susceptibility. The requirement of overlapping histograms
implies that $\delta m$ satisfies
\begin{equation}
\delta m \sim 1 /\sqrt{\langle\chi\rangle V} \sim m_{\sigma}/\sqrt{V}.
\end{equation}

The overhead depends on the step size $\delta m$.  As the susceptibility is
related to the scalar meson mass, $\chi = A_{\sigma}/m_{\sigma}^{2}$, the step
size is large in the chiral limit. Hence simulated tempering becomes more
effective for very small quark masses, with the gain in speed more than
compensating for the $N_{m}^{2}$ cost of having additional masses.

The volume dependence above is in units of some relevant
correlation length, so taking the continuum limit in fixed physical volume does
not cause the method to break down.

A further improvement comes if you temper in a parallel way.
If  $N_{m}$ different masses are needed, you can happily do $N_{m}$ different
simulated tempering runs simultaneously on different computers. You can even go
one better, and put them together on one computer in a way that removes the
need for the constants $g_{i}$ and improves the performance. This is called
parallel tempering; results are presented in the next section.

In parallel tempering you first generate one (thermalised) configuration
$C_{i}$ at each of the masses.  Then use a Metropolis step to decide whether
the configuration $C_{i}$ at mass $m_{i}$, and configuration $C_{i+1}$ at mass
$m_{i+1}$ should be swapped for the next trajectory. If this is done, the next
trajectory will run with $C_{i+1}$ at mass $m_{i}$, and $C_{i}$ at mass
$m_{i+1}$. After each trajectory one starts trying to swap masses 1 and 2,
moving up through the list, ending by trying to swap the configurations at
masses $N_{m}-1$ and $N_{m}$.

This method doesn't need any constants $g_{i}$, changes the mass at two rather
than one configuration, and has the further advantage of generating a
configuration at each mass every trajectory. Also, trajectories way out in the
tail of the distribution stand a chance of moving more than one step in
mass. This doesn't happen very often though!

\section{RESULTS}

\begin{table}
\begin{center}
\caption{Parameters and results from the tempered (T) and a standard HMC
run. The trajectories have unit length.  The last three columns
give the acceptance rate for the Metropolis mass changes, and the 
integrated autocorrelation times of the plaquette
and $2\times 2$ Wilson loop.}
\label{tab:tauint}
\vspace{\baselineskip}
\begin{tabular}{r@{\extracolsep{\fill}}rrrl}
\hline\hline
\multicolumn{1}{c}{$m$}&
\multicolumn{1}{c}{s/trj}&
\multicolumn{1}{c}{\%}&
\multicolumn{1}{c}{$\tau_{\text{int}}^{P}$}&
\multicolumn{1}{c}{$\tau_{\text{int}}^{W2\times 2}$}
\\
\hline
(T)   0.020 &  475  & 15 & 4.9(14) & 3.6(10) \\ 
(T)   0.024 &  350  & 21 & 3.3(8) & 2.8(6)  \\ 
(T)   0.028 &  283  & 24 & 4.2(15) & 3.3(10) \\ 
(T)   0.032 &  231  & 21 & 3.5(9) & 2.7(10) \\ 
(T)   0.036 &  193  & -- & 3.6(12) & 2.8(6)  \\ 
(HMC) 0.020 &  475  & -- & 9.0(21)& 8.4(17) \\ 
(HMC) 0.040 &  166  & -- & 8.5(20) & 6.9(15) \\ 
(HMC) 0.060 &   84  & -- & 7.3(13) & 5.8(13) \\ 
\hline\hline 
\end{tabular}
\end{center}
\end{table}

Two sets of runs are in progress with four staggered fermions on an
$8^{3}\times 12$ lattice. The tempered one has five masses, $\{0.020, 0.024,
0.028, 0.032, 0.036 \}$, called set `T'. For comparison there are three
standard HMC runs at masses 0.020, 0.040 and 0.060. The run parameters are
given in table~\ref{tab:tauint}. So far about 2000 units in $\tau$ have been
run for T, and about 3500 for the HMC run.

\begin{figure}[t]
    \begin{center}
      \leavevmode
     \epsfig{file=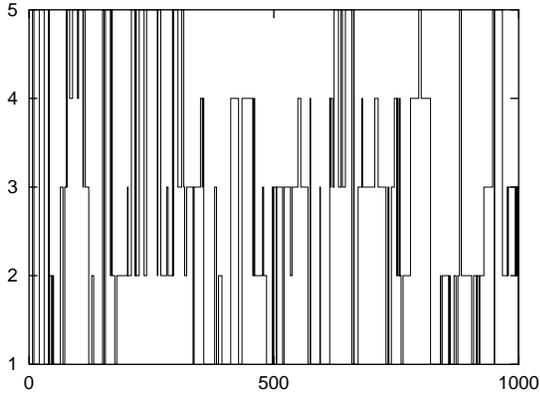}
      \caption{Part of the time history of the configuration number running at
 	$m=0.020$ in the tempered run.
        }
      \label{fig:mass}
    \end{center}
\end{figure}

Which of the five configurations of the tempered run used $m=0.020$ at which
time can be seen in the time history of figure~\ref{fig:mass}. It is clear that
all five configurations have run at each mass about equally often, as required.

The acceptance rate for transitions between masses is also shown in
table~\ref{tab:tauint}, and lies around 20\%. Another run with closer masses,
$\{0.020, 0.023, 0.026, 0.029, 0.032 \}$, yielded rates about ten percentage
points higher. An acceptance rate of around 20\% to 30\% seems optimal.

A measure of the speed of an algorithm is $\tau_{\text{int}}^{O}$, the
integrated auto-correlation~\cite{tauint} for an observable $O$. With
insufficient data, $N_{\text{data}}<1000\tau_{\text{int}}$, an accurate value
cannot be obtained.
The largest value for $\tau_{\text{int}}^{O}$ of all observables
$O$ defines the number of independent configurations.

For full QCD the global topological charge $Q$ seems to be the slowest
observable\cite{us} . However, on this size lattice the topology 
(field theoretic definition) turned out to depend on the action
used for cooling, and is probably not well defined\footnote{The cooling actions
tested used $1\times 1$ and $1\times 2$ loops with various values of the
coefficients. On a given configuration different choices for the action lead to
completely different global topological charges.}.  
The plaquette and Wilson
loops up to $2\times 2$ seem to have the most well defined auto-correlation,
and have been used here.

\begin{figure}[t]
    \begin{center}
      \leavevmode
     \epsfig{
file=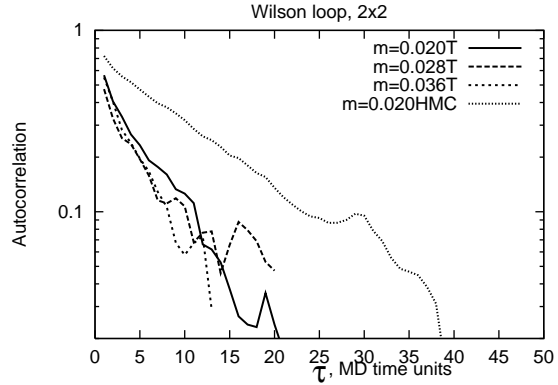}
      \caption{The auto-correlation function for the $2\times 2$ Wilson loop at 
	$m=0.020$ from set T and a standard HMC reference run.
        }
      \label{fig:cvect}
    \end{center}
\end{figure}

In figure~\ref{fig:cvect} the auto-correlation function for the $2\times 2$
Wilson loop at $m=0.020$ from the tempering and a standard HMC run is
plotted. The integrated autocorrelations obtained for both the plaquette and
the $2\times 2$ Wilson loop are given in table~\ref{tab:tauint}. These turn out
to be about 20\% smaller than the slope parameter needed to fit the central
part of the correlator in figure~\ref{fig:cvect} to an exponential.

For observables from the tempered run, $\tau_{\text{int}}$ is about three
times smaller than from the standard HMC run. The computer time needed
per tempered trajectory, $T_{\text{T}}$, compared with the time for a single
HMC run at the smallest mass yields
$T_{\text{T}}= 3.22 T^{m=0.02}_{\text{HMC}}$.

%

\section{CONCLUSIONS}
Tempering yields an integrated auto-correlation that is about a factor of 
three smaller than the HMC run, although much better data is needed to make
this reliable! Tempering costs about three times more than the single 
HMC run at the smallest mass, so it is clearly faster if more than one
mass is required, as is usually the case!

For realistic simulations, using improved actions on large, smooth lattices at
very small quark masses, tempered methods are very likely to be of
benefit. This is especially true for Wilson fermions, where many $\kappa$
values are needed in order to extract meaningful physics.

\section*{Acknowledgements}
This work was partially supported by HCM-Fellowship contract
ERBCHBGCT940665, and currently the Japanese Society for the 
Promotion of Science. The author is grateful to both the Pisa and Tsukuba
lattice groups for many useful discussions related to this work.


\begin{thebibliography}{10}

\bibitem{us} G. Boyd {\em et al.},
Proceedings, Lattice 96, IFUP-TH 47/96, hep-lat/96008123;
B. All\'{e}s, {\em et al.}, Phys. Lett {\bf B 389} (1996) 107;
K. Bitar {\em et al.}, Phys. Rev. {\bf D42} (1990) 3794.

\bibitem{ST} E.~Marinari and G.~Parisi,
Europhys. Lett. {\bf 19} (1992) 451;
A.~P.~Lyubartsev {\em et al.},
J. Chem. Phys. {\bf 96} (1992) 1776;
L. Fern\'andez {\em et al.},
J. Phys. I France {\bf 5} (1995) 1247; E. Marinari, proceedings,
Bielefeld workshop on Multi-scale Phenomena, October 1996.
and cond-mat/9612010.

\bibitem{st_biel} G. Boyd, proceedings,
Bielefeld workshop on Multi-scale Phenomena, October 1996; hep-lat/9701009.

\bibitem{hmca} 
S.~Gottlieb {\em et al.}, Phys. Rev. {\bf D 35}
 (1987) 2531; A.~D.~Kennedy, Nucl.\ Phys.\ {\bf B}
(Proc. Suppl.) {\bf 30} (1993) 96.

\bibitem{tauint} See, for example, A. D. Sokal, in {\em Quantum Fields on the
Computer}, Ed. M. Creutz, World Scientific 1992.


\end{thebibliography}
\end{document}